\documentclass[a4paper,11pt]{article}
\usepackage{pos}

\usepackage[sort&compress]{natbib}
\usepackage{xcolor}
\usepackage{ulem}

\newcommand{\be}{\begin{equation}}  
\newcommand{\ee}{\end{equation}}  
\newcommand{\beq}{\begin{eqnarray}} 
\newcommand{\eeq}{\end{eqnarray}}

\title{First study of twist-3 PDFs for the proton from lattice QCD}
\author[a]{S. Bhattacharya}
\author[b]{K. Cichy}
\author*[a]{M. Constantinou}
\author[a]{A. Metz}
\author[a]{A. Scapellato}
\author[c]{F. Steffens}

\affiliation[a]{Department of Physics,  Temple University,  \\ Philadelphia,  PA 19122 - 1801,  USA}
\affiliation[b]{Faculty of Physics, Adam Mickiewicz University, \\ ul.\ Uniwersytetu Pozna\'nskiego 2, 61-614 Pozna\'{n}, Poland}
\affiliation[c]{Institut f\"ur Strahlen- und Kernphysik, Rheinische
  Friedrich-Wilhelms-Universit\"at Bonn, \\ Nussallee 14-16, 53115 Bonn}

\emailAdd{marthac@temple.edu}

\abstract{In these proceedings, we summarize the main results from the first-ever calculations of the chiral-even and chiral-odd twist-3 parton distributions, $g_T(x)$ and $h_L(x)$, of the proton from lattice QCD. 
We use an $N_f=2+1+1$ ensemble of maximally twisted mass fermions with a clover improvement. 
The lattice has a spatial extent of 3~fm, the lattice spacing is 0.093~fm, and the pion mass is $260$~MeV. The matrix elements are obtained with a source-sink time separation of 1.12~fm to control contamination from excited states. The calculation is based on the quasi-PDF approach and employs three values for the proton momentum: 0.83~GeV, 1.25~GeV, and 1.67~GeV. 
The lattice data are renormalized non-perturbatively using the RI$'$ scheme, and the final results are presented in the $\overline{\rm MS}$ scheme at the scale of 2~GeV. Furthermore, we compute in the same setup the helicity, $g_1(x)$, and transversity, $h_1(x)$, distributions, which are used to compare $g_T(x)$ and $h_L(x)$ to their Wandzura-Wilczek approximations. For $h_L(x)$, we combine results for the isovector and isoscalar flavor combinations to disentangle the individual up- and down-quark contributions.}

\FullConference{%
 The 38th International Symposium on Lattice Field Theory, LATTICE2021
  26th-30th July, 2021
  Zoom/Gather@Massachusetts Institute of Technology
}

\begin{document}
\maketitle

\section{Introduction}

Parton distribution functions (PDFs) are essential in characterizing the structure of hadrons in terms of their constituent partons within quantum chromodynamics (QCD). PDFs and their generalizations can be obtained from cross sections of high-energy processes through factorization theorems of QCD. PDFs are characterized by their twist, which indicates the order in the large scale of the process, $1/Q$, at which they appear in the cross section~\cite{Jaffe:1996zw}. While the leading-twist PDFs (twist-2) have been extensively studied both experimentally and theoretically, twist-3 PDFs are poorly known. This poses limitations on our understanding of hadron structure, because twist-3 PDFs are important too. Their significance is partly because they can be of similar magnitude as their twist-2 counterparts. Also, they contain information about quark-gluon correlations inside the hadrons~\cite{Balitsky:1987bk}, and some twist-3 PDFs are connected to  transverse momentum dependent parton distributions~\cite{Accardi:2009au, Gamberg:2017jha, Cammarota:2020qcw}. It should be noted that twist-3 PDFs can have physical meaning, even though they lack probabilistic interpretation~\cite{Burkardt:2008ps}. Measurements related to twist-3 PDFs are part of the 12 GeV program at Jefferson Lab and are important for the Electron-Ion Collider~\cite{AbdulKhalek:2021gbh}. However, measuring twist-3 PDFs is difficult, as they are suppressed by the ${\mathcal{O}(1/Q)}$ kinematic factor. Therefore, calculations from first principles are highly valuable. Recent reviews on calculations of $x$-dependent distribution functions can be found in Refs.~\cite{Cichy:2018mum,Ji:2020ect,Constantinou:2020pek,Cichy:2021lih}.

Here, we present the first-ever calculation of the twist-3 PDFs $g_T(x)$ and $h_L(x)$, within lattice QCD. $g_T(x)$ is relevant for the polarized structure functions $g_1^{\rm s.f.}$ and $g_2^{\rm s.f.}$, which enter the cross section of the DIS process. The chiral-odd PDF $h_L(x)$ is not related to the ``simple'' DIS process, and therefore, experimental experimental information is almost non-existing. It could be measured through the double-polarized Drell-Yan process~\cite{Jaffe:1991kp,Jaffe:1991ra} or single-inclusive particle production in proton-proton collisions~\cite{Koike:2016ura}. For $g_T$, we focus on the isovector flavor combination that receives contributions from the connected diagram only. For $h_L(x)$, we perform a flavor decomposition, because the contribution from the disconnected diagram is expected to be small~\cite{Alexandrou:2021oih}.

\section{Lattice calculation}

We employ the quasi-PDF formalism to extract the twist-3 PDFs for the proton, which requires the calculation of matrix elements containing non-local matrix elements,
\begin{equation}
{\mathcal M}_{\cal O}(z,P)\,=\,\langle P \,\vert\, \overline{\psi}(0,\vec{0})\,{\cal O}\, W(0,\vec{z})\,\psi(0,\vec{z})\,\vert P\rangle\,.
    %{\mathcal M}_{h_L}(P,z)\,=\,\langle P\vert \, \overline{\psi}(0,z)\,\sigma_{jk}\, \tau_3 W(z)\,\psi(0,0)\,\vert P\rangle\,.
\label{eq:matrix_elements}
\end{equation}
$\vert P\rangle$ denotes a proton state with four-momentum $P=(iE,0,0,P_3)$. For gauge invariance, a straight Wilson line is inserted between the fermion fields of the operator. The fermion fields, $\psi$ and $\bar{\psi}$, are separated by a space-like distance $z$. To relate the matrix element to the physical PDFs, the Wilson line is in the direction of the boost, $P_3$. The operators we study are $\gamma^5\,\gamma^j$ and $\sigma_{jk}$ ($j,\,k=1,2$, $j\ne k$), that correspond to $g_T(x)$ and $h_L(x)$, respectively. The ground-state of ${\mathcal M}_{\cal O}$ is related to the function $F_{\cal O}$, which represents the quasi-PDFs in coordinate space and are, thus, $z$-dependent,
\beq
F_{g_T}(z,P_3,\mu) &=& - i \frac{E}{m} Z_{\gamma^5\,\gamma^j}(z,\mu)\, {\cal M}_{\gamma^5\,\gamma^j}(P_3,z)\,,\\[0.5ex]
F_{h_L}(z,P_3\mu) &=& -i\,\epsilon_{jk30}\,\frac{E}{m} Z_{\sigma_{jk}}(z,\mu)\, M_{\sigma_{jk}}(z,P_3)\,.
\eeq
In these equations, $m$ is the proton mass, and $E{=}\sqrt{m^2+P_3^2}$ is the energy of the state with momentum $P_3$. $Z_{\cal O}$ is the renormalization function of the operator, which we calculate non-perturbatively in the RI' scheme and convert to the modified $\overline{\rm MS}$ scheme at a scale $\mu=2$ GeV. More details can be found in Refs.~\cite{Bhattacharya:2020cen,Bhattacharya:2021moj}. 

A Fourier transform is applied to the functions $F_{\cal O}(P_3,z)$ to form the quasi-PDFs, $\widetilde{q}$, in the momentum representation (square brackets in Eq.~\eqref{eq:FT}). Finally, these are matched to their light-cone counterparts using a perturbative formula obtained within Large Momentum Effective Theory (LaMET) \cite{Ji:2013dva,Ji:2014gla},
\begin{equation}
\label{eq:FT}
q_{\cal O}(x,\mu)=\int_{-\infty}^\infty 
\frac{d\xi}{|\xi|} \, C_{\cal O}\left(\xi,\frac{\mu}{x P_3}\right)\,
\left[2 P_3 \int_{-\infty}^{+\infty}\hspace*{-0.1cm}\frac{dz}{4\pi}\,
e^{-i\frac{x}{\xi}P_3z}\,{ F}_{\cal O}(P_3,z) \right]\,.
\end{equation}

\noindent
In practice, instead of the naive Fourier transform, we use the Backus-Gilbert method, which provides a unique solution for the quasi-PDFs. $C_{\cal O}$ is the matching kernel, which we calculated within one-loop perturbation theory in momentum space~\cite{Bhattacharya:2020xlt,Bhattacharya:2020jfj}. It should be noted that the matching formalism does not take into account mixing with quark-gluon-quark operators, which would also require matrix elements of the latter. An alternative matching formalism that discusses such a mixing can be found in Refs.~\cite{Braun:2021gvv,Braun:2021aon}. The matching procedure for twist-3 is much more complicated than for twist-2, due to the presence of zero-modes that require special attention. The matching expression for $C_{\cal O}$ connects the quasi-PDFs in the modified $\overline{\rm MS}$ scheme with the light-cone PDFs in the $\overline{\rm MS}$ scheme at 2 GeV. 

%\bigskip
The calculation is performed on an $N_f=2+1+1$ ensemble of twisted mass fermions with clover improvement, produced by the ETM collaboration (ETMC)~\cite{Alexandrou:2021gqw}. The pion mass of the ensemble is 260 MeV, the lattice spacing is $a\simeq 0.093$ fm and the lattice volume is $32^3\times 64$ ($L\approx 3$ fm). We produce the matrix elements for both operators at the boosts $P_3=0.83,\, 1.25,\, 1.67$ GeV with 1552, 11696, and 105216 measurements, respectively.

\section{Results}

\begin{figure}[h!]
\begin{minipage}{5cm}
\includegraphics[scale=0.55]{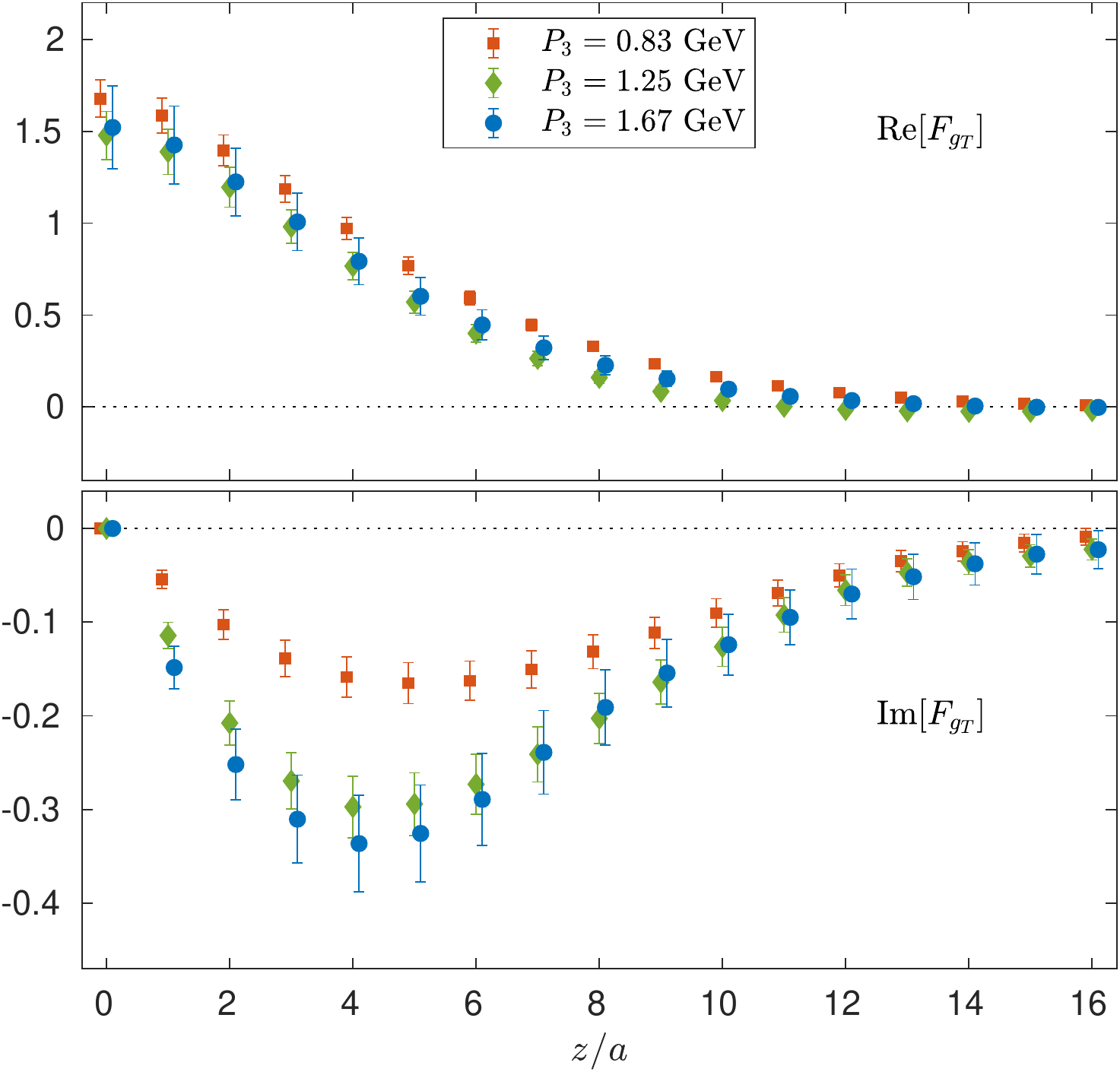}
\end{minipage}    
\hfill
\begin{minipage}{6.5cm}
\includegraphics[scale=0.415]{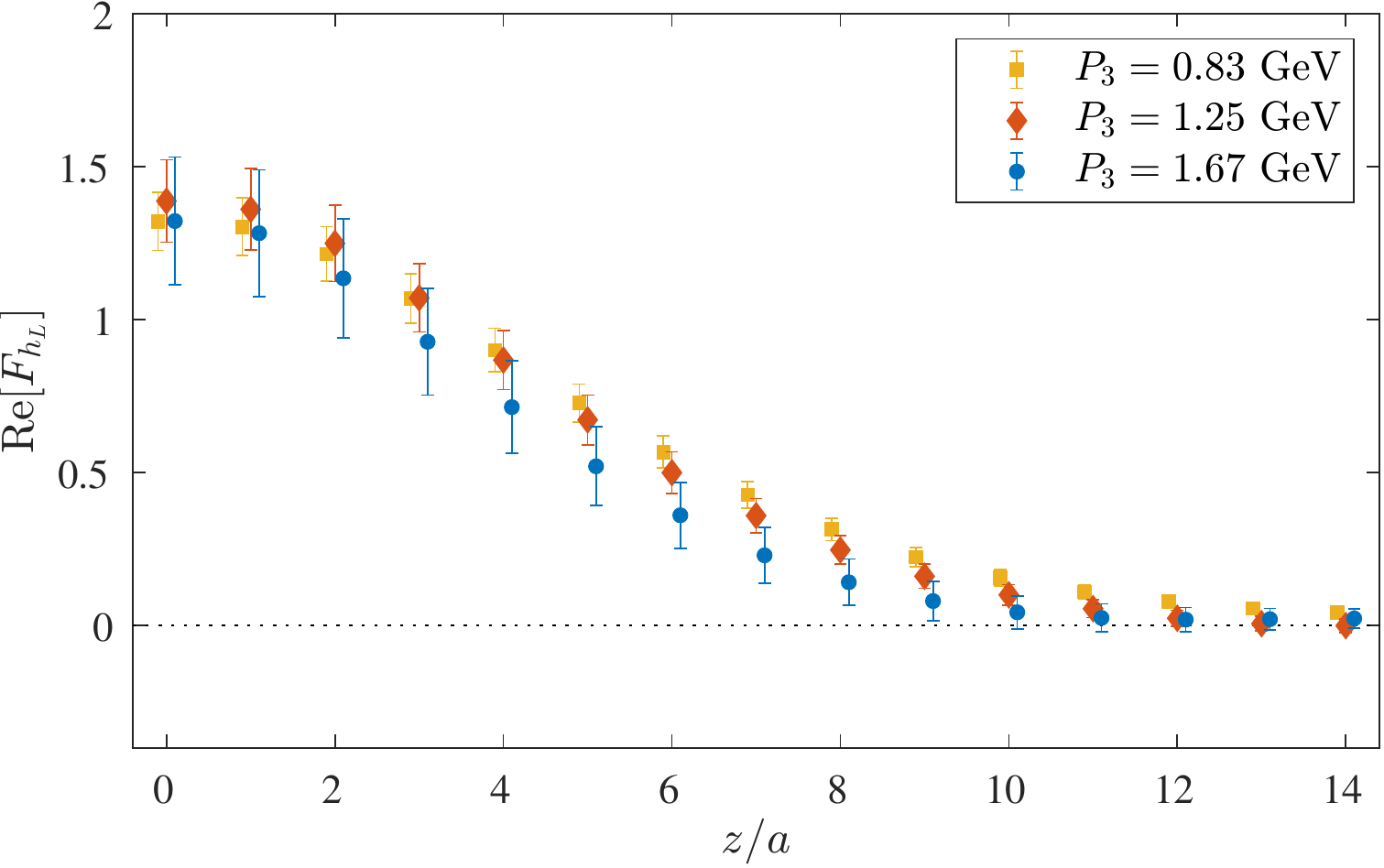}\\
\includegraphics[scale=0.415]{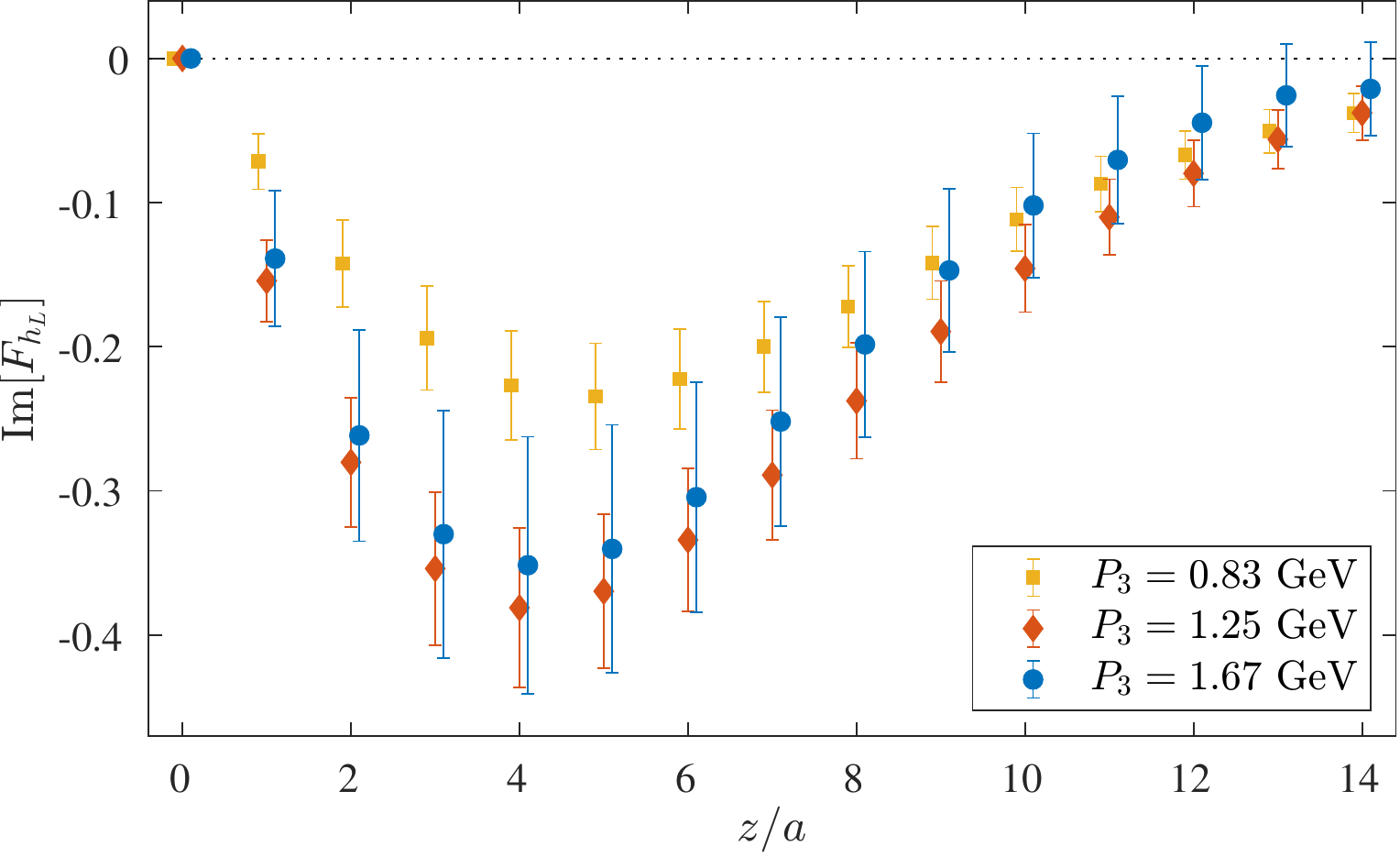}
\end{minipage}   
\vspace*{-0.3cm}
 \caption{Left: Real (top) and imaginary (bottom) parts of ${F}_{g_T}$ for $P_3=0.83,\,1.25,\,1.67$ GeV, shown with red squares, green diamonds, blue circles, respectively. Right: Real (left) and imaginary (right) part of $F_{h_L}$ for momenta $0.83$~GeV (yellow squares), $1.25$~GeV (red diamonds) and $1.67$~GeV (blue circles).}
 	\label{fig:Fs}
\end{figure}

In Fig.~\ref{fig:Fs} we show the functions $F_{g_T}$ and $F_{h_L}$ in coordinate space for $P_3=0.83,\,1.25,\,1.67$ GeV. In all cases, we find convergence between the two largest momenta within statistical uncertainties. The matching formula depends on $P_3$, and therefore, such a convergence is not guaranteeed in the final PDFs. In our calculations, we find that a convergence between $P_3=1.25$ GeV and $P_3=1.67$ GeV holds for both $g_T(x)$ and $h_L(x)$.

\begin{figure}[h!]
  \hspace*{-0.2cm} 
  \includegraphics[scale=0.52]{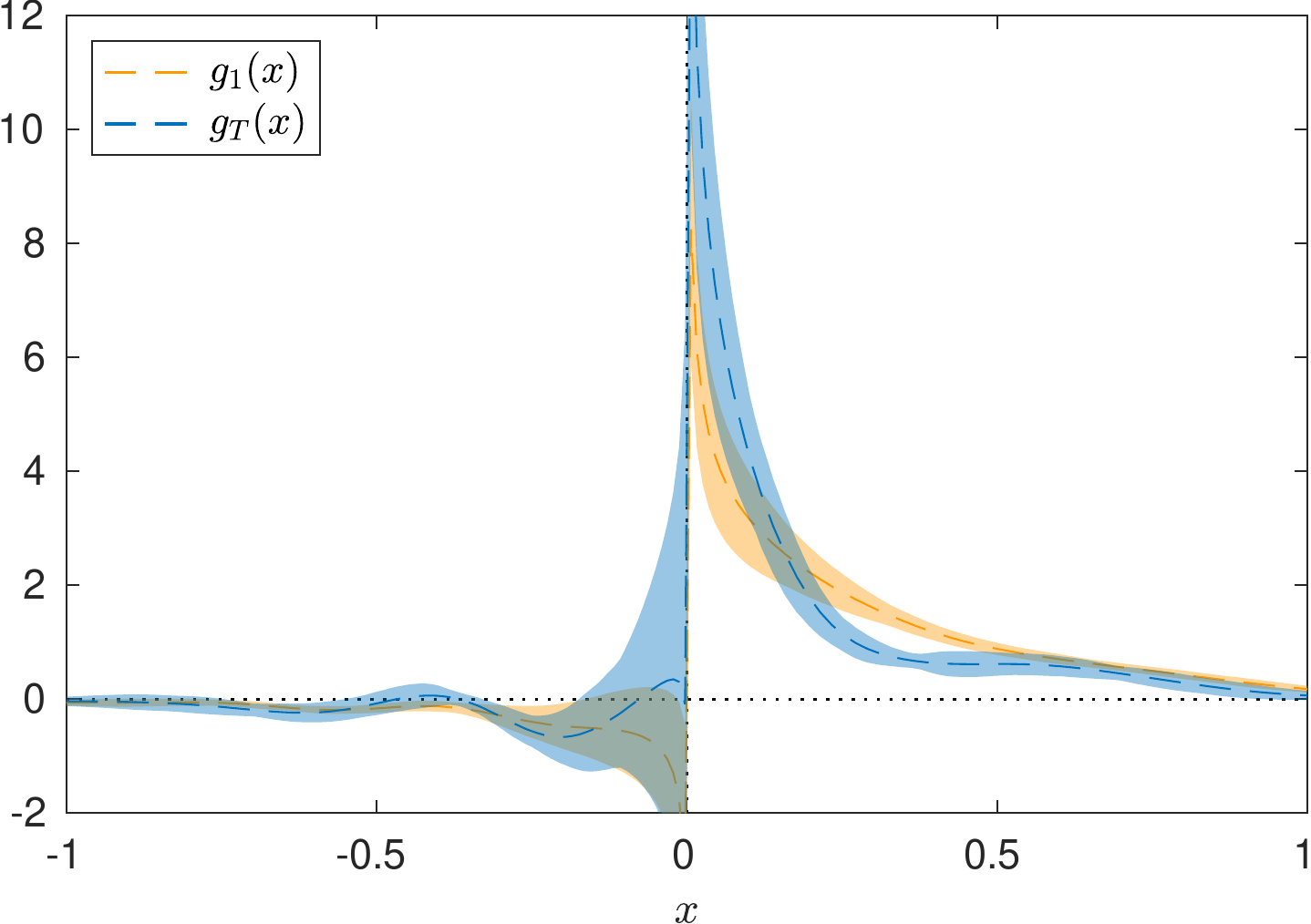}
  \includegraphics[scale=0.52]{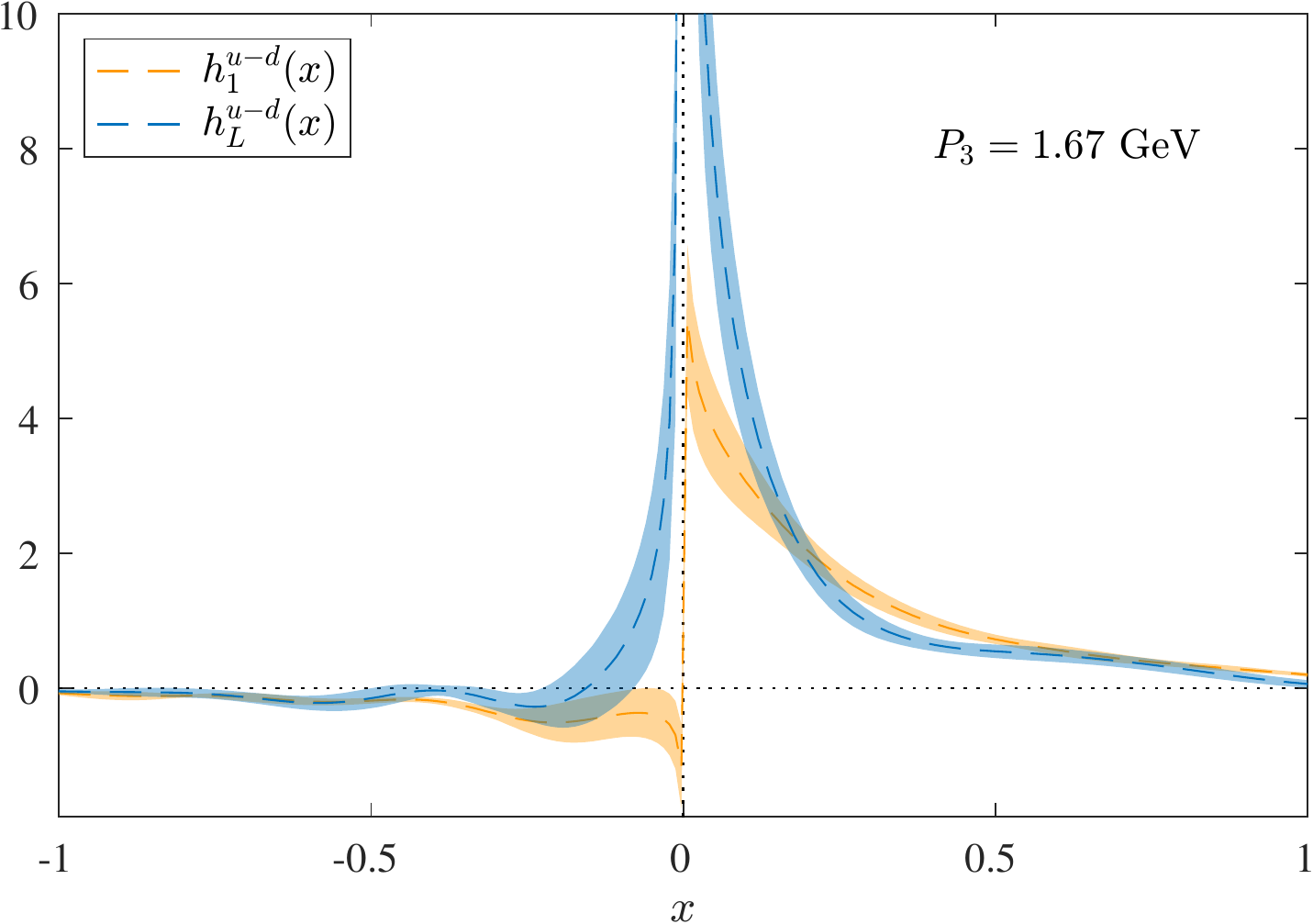}
\vspace*{-0.75cm}
 	\caption{Left: Comparison of $g_T$ (blue band) and $g_1$ (orange band) at $P_3=1.67$ GeV. Right: Comparison of $h_L$ (blue band) and $h_1$ (orange band) for the nucleon boost $P_3=1.67$~GeV.}
 	\label{fig:gT_g1_hL_h1}
 \end{figure}

We also obtain the twist-2 counterparts of $g_T(x)$ and $h_L(x)$, that is $g_1(x)$ and $h_1(x)$. These are used to compare them with the twist-3 PDFs and check the significance of the latter. Such a comparison can be found in Fig.~\ref{fig:gT_g1_hL_h1} for the highest momentum, $P_3=1.67$ GeV. Comparing $g_1(x)$ and $g_T(x)$, we find that they differ in the positive-$x$ region. In particular, $g_1(x)$ is smaller than $g_T(x)$ in the low-$x$ region and has a smaller slope at $x\approx0.1-0.3$. The two distributions are in agreement in the antiquark region within uncertainties and the large positive-$x$ region. A similar situation is observed for $h_L(x)$, which is as sizeable as its twist-2 counterpart, $h_1(x)$. We emphasize that the PDFs extracted from lattice QCD are not reliable in the region $|x| \lesssim 0.15$ mainly due to higher-twist effects. For $x > 0.15$, $h_1(x)$ is dominant only for $0.2 \lesssim x \lesssim 0.5$. $h_1(x)$ and $h_L(x)$ are in agreement in the anti-quark region for $x<-0.15$. The bands include statistical uncertainties, as well as systematic ones due to the $z_{\rm max}$ that enters the reconstruction of the $x$ dependence.

\section{Burkhardt-Cottingham sum rules}

The integrals of twist-3 PDFs and their twist-2 counterparts are connected through the Burkhardt-Cottingham (BC) sum rule~\cite{Burkhardt:1970ti,Burkardt:1995ts}, which have also been extended to the quasi-PDFs~\cite{Bhattacharya:2021boh}. These sum rules are very useful for the qualitative understanding of twist-3 distributions, and it is interesting to test these relations using our lattice data. For instance, the integrals of $g_1(x)$ and $g_T(x)$ should coincide, that is,
\vspace*{-0.1cm}
\begin{equation}
\label{eq:SR}
\int_{-1}^{1} dx\,g_1(x) - \int_{-1}^{1} dx\,g_T(x) = 0\,.
\end{equation}
Here we find a value of 0.01(20), and therefore the sum rule is satisfied, which suggests that effects due to operator mixing could be relatively small. There are a number of tests to perform on $h_1(x)$ and $h_L(x)$, such as the sum rule for their quasi-PDFs, $\widetilde{h}_1(x)$ and $\widetilde{h}_L(x)$,
\begin{equation}
\int^1_{-1} dx \,\tilde{h}_L(x,P_3) = \int^1_{-1} dx \,\tilde{h}_1(x,P_3) = g^T\,.
\end{equation} 
The tensor charge $g^T$ is independent of the kinematic setup, and the relation should hold for all $P_3$. We tested this equality numerically, and here we show the results for the highest momentum,
\begin{eqnarray}
\label{eq:BC1}
\int dx\, \widetilde{h}_L(x, 1.67\,{\rm GeV})=1.03(16)\,,&\quad&  \int dx\, \widetilde{h}_1(x, 1.67\,{\rm GeV})=0.94(10) \,,
\label{eq:BC3}
\end{eqnarray}
for which we find agreement in the extracted value of the tensor charge. In fact, the sum rule is satisfied at each momentum, and is independent of the momentum, as indicated in Ref.~\cite{Bhattacharya:2021boh}. 

\section{Wandzura-Wilczek approximation}

The twist-3 $g_T(x)$ and its corresponding twist-2 $g_1(x)$ are connected, at a given $x$, through the Wandzura-Wilczek (WW) approximation~\cite{Wandzura:1977qf}. An analogous relation exists for $h_L(x)$~\cite{Jaffe:1991ra,Jaffe:1991kp}, in which the Mellin moments of $h_L(x)$ can be split into twist-2 and twist-3 parts.
\begin{eqnarray}
\label{eq:gT_WW}
g_T(x) &=& g_T^{\rm WW}(x) + g^{\rm twist-3}_T(x) = \int_x^1 \frac{dy}{y} g_1(y) + g^{\rm twist-3}_T(x)\,, \\[1.25ex]
    h_L(x) &=& h_L^{\rm WW}(x) + h_L^{\rm twist-3}(x) 
    =  2x\int_x^1 dy \, \frac{h_1(y)}{y^2} + h_L^{\rm twist-3}(x) \,.
    \label{e:WW}
\end{eqnarray}
In the WW approximation, $g_T(x)$ and $h_L(x)$ are fully determined by the twist-2 $g_1(x)$ and  $h_1(x)$, respectively. $g_T^{\rm twist-3}(x)$ and $h_L^{\rm twist-3}(x)$ are genuine twist-3 contribution, which is given by quark-gluon correlations (and a current-quark mass term). In the WW approximation one only keeps the term $g_T^{\rm WW}(x)$ ($h_L^{\rm WW}(x)$), which is determined by the helicity (transversity) distribution.

\begin{figure}[h!]
\hspace*{-0.35cm}
\includegraphics[scale=0.54]{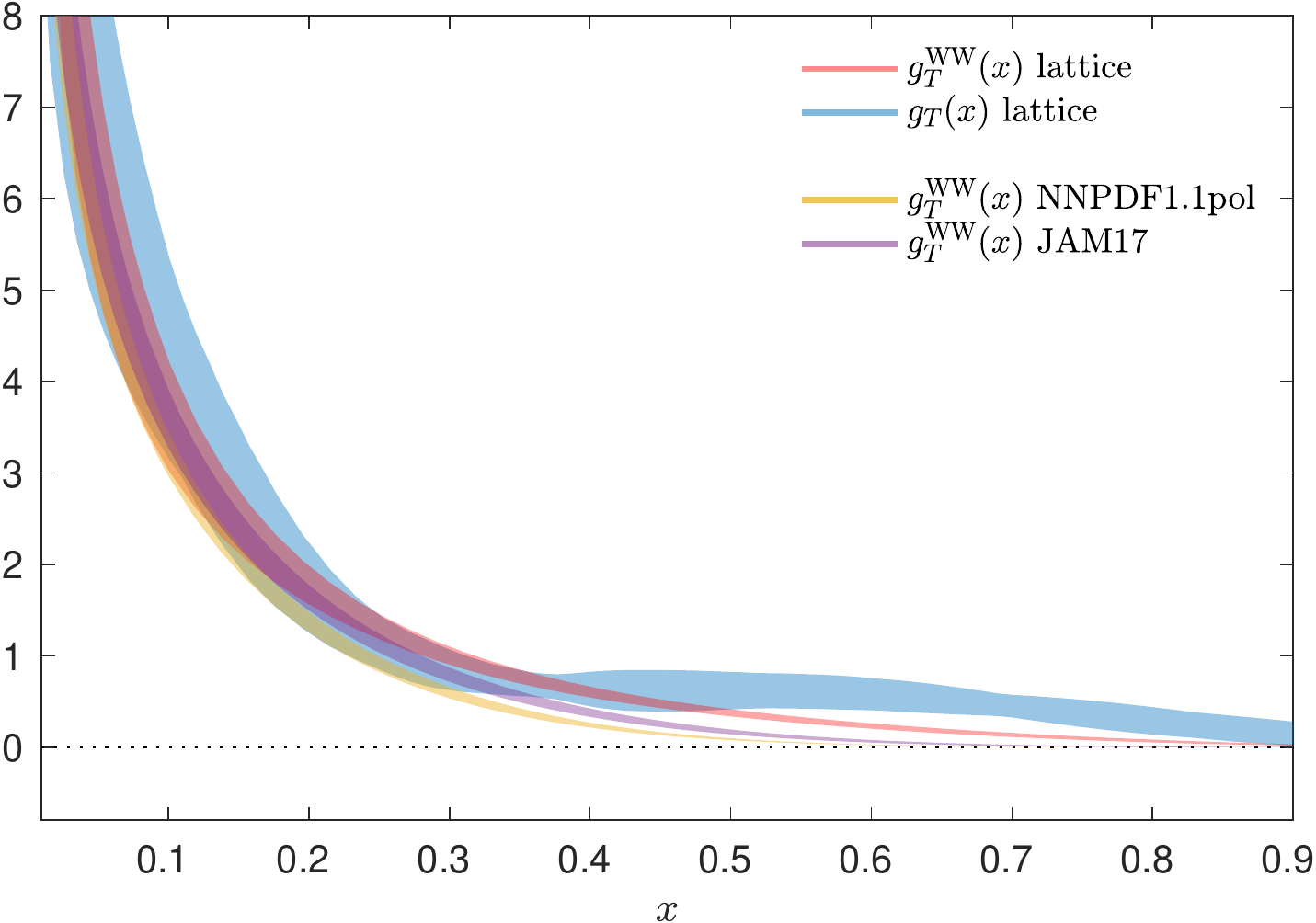}
    \includegraphics[scale=0.54]{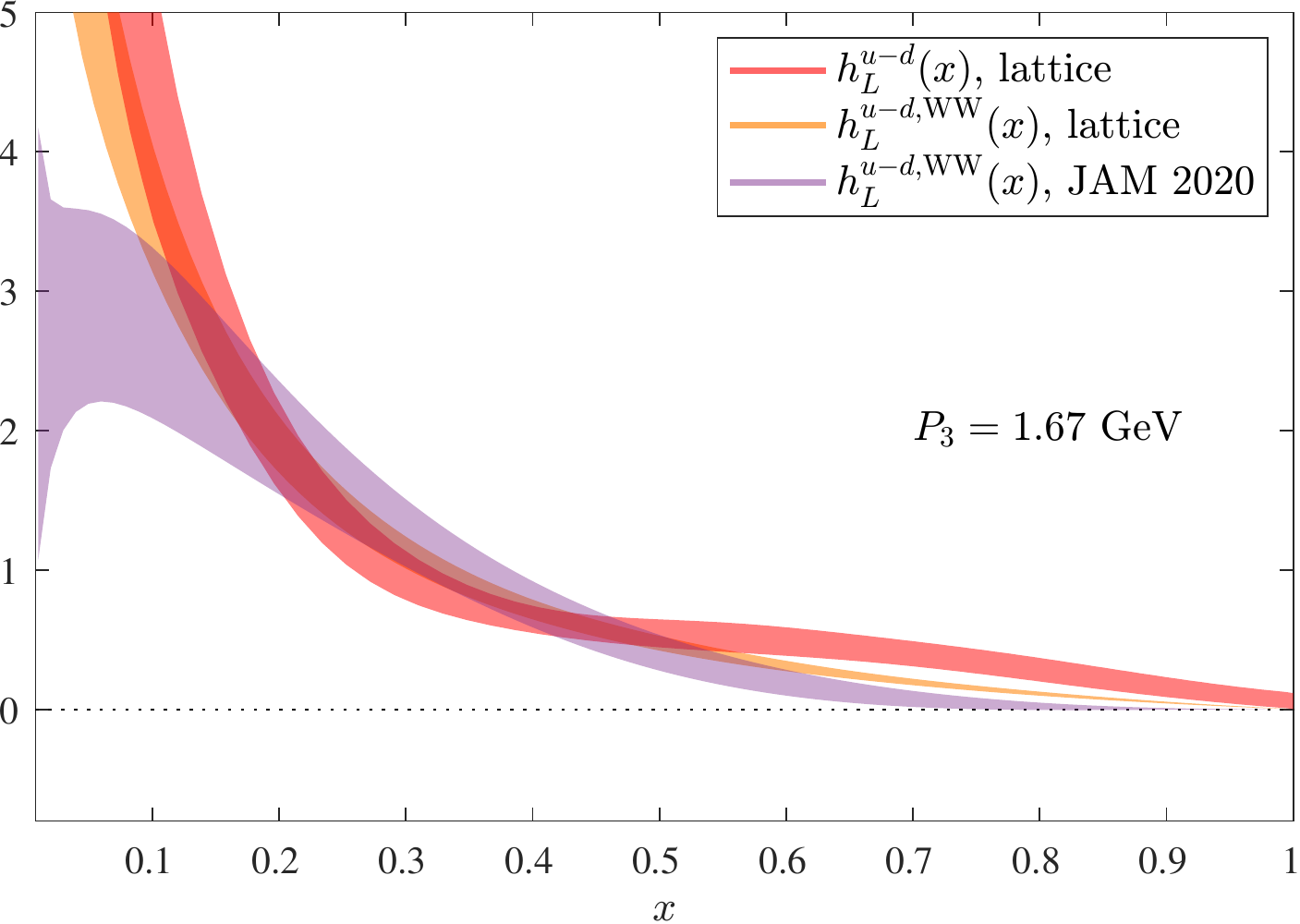}
    \vspace*{-0.3cm}
	\caption{Left: Comparison of our $g_T(x)$ at $P_3=1.67$ GeV (blue band) with its WW estimates: lattice-extracted $g_T^{\rm WW}$ (red band) and global fits-extracted (NNPDF1.1pol~\cite{Nocera:2014gqa} orange band, JAM17~\cite{Ethier:2017zbq} purple band). Right: The WW approximation for $h_L(x)$, for boosts $P_3=1.67$ GeV. The lattice estimate of $h_L(x)$ (red band) is compared with its WW-approximation (orange band) extracted on the same gauge ensemble and the one obtained from global fits (violet band) from the JAM collaboration~\cite{Cammarota:2020qcw}.}
 	\label{fig:WW_approx}
 \end{figure}

The check of the WW approximation for $g_T(x)$ and $h_L(x)$ is another highlight of our calculation. This is important, because, for example, it was found in the instanton model of the QCD vacuum that the lowest nontrivial moment of $h_L^{\rm twist-3}(x)$ is numerically very 
small~\cite{Dressler:1999hc}. Using our results, we can also provide qualitative conclusions on the significance of the contribution due to quark-gluon correlations in Eq.~(\ref{e:WW}).
The WW approximation for both $g_T(x)$ and $h_L(x)$ is shown in Fig.~\ref{fig:WW_approx} for $P_3{=}1.67$ GeV. We only show the quark region ($x{>}0$), which is less susceptible to systematic uncertainties than the antiquark region. We see that the actual lattice data for $g_T(x)$ are consistent with $g_T^{\rm WW}(x)$ for a considerable $x$-range. However, given the uncertainties, a violation of the WW approximation is still possible at the level of up to 40\% for $x{\lesssim}0.4$. This is consistent with the findings of Ref.~\cite{Accardi:2009au} based on experimental data, estimating a possible violation of the WW approximation of ${\sim}15{-}40\%$. We also find that the slopes of $g_T$ and $g_T^{\rm WW}$ are the same up to $x{\approx}0.4$. The difference of $g_T$ and $g_T^{\rm WW}$ for large $x$ could be either due to unquantified systematic uncertainties, or due to larger violations of the WW approximation in this region. Our results on $g_T^{\rm WW}$ are compared to the estimate obtained using $g_1$ from global fits by the NNPDF \cite{Nocera:2014gqa} and JAM \cite{Ethier:2017zbq} collaborations; we find good agreement up to $x{\approx}0.3$.
The results of $h_L(x)$ and $h^{\rm WW}_L(x)$ are in agreement for $x{\lesssim}0.55$. Our lattice results on $h^{\rm WW}_L(x)$ in the region $0.15 \lesssim x{\lesssim}0.55$ are also in good agreement with $h^{\rm WW}_L(x)$ obtained from a global fit of the nucleon transversity by the JAM collaboration~\cite{Cammarota:2020qcw}. These findings seem to suggest that $h_L(x)$ could be determined by the twist-2 $h_1(x)$ for a considerable $x$-range. More precise statements require further investigations. We repeat that the mixing with quark-gluon-quark operators has not been computed within this work, and other systematic effects have to be addressed as well, like those related to a finite lattice spacing and a non-physical light quark mass. In addition, the tension observed between global fits and lattice data at small and large $x$ reveals that more control may be needed to constrain distributions in these regions.

\begin{figure}[h!]
    \includegraphics[scale=0.52]{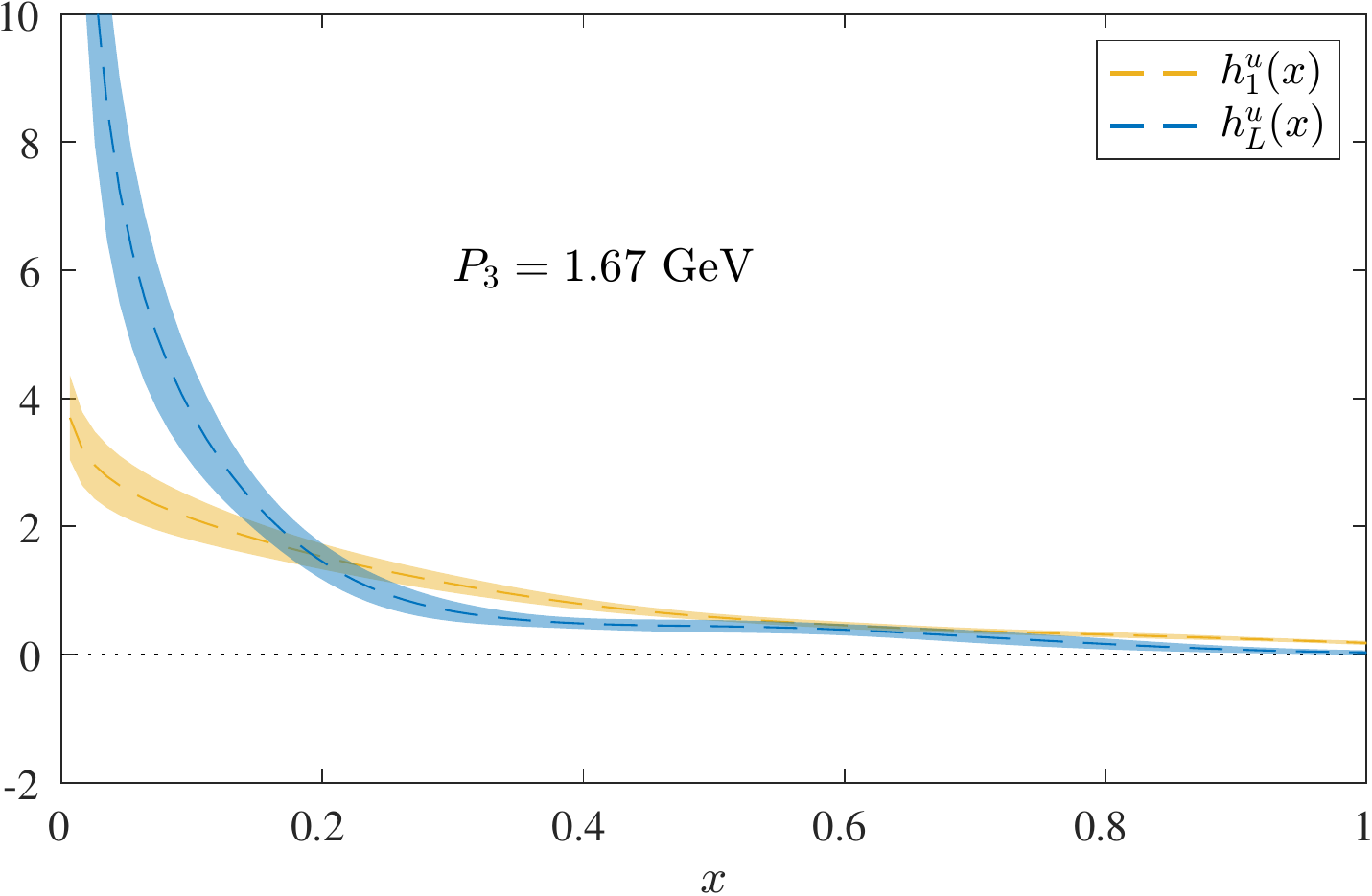}\hspace{0.1cm}
   \includegraphics[scale=0.52]{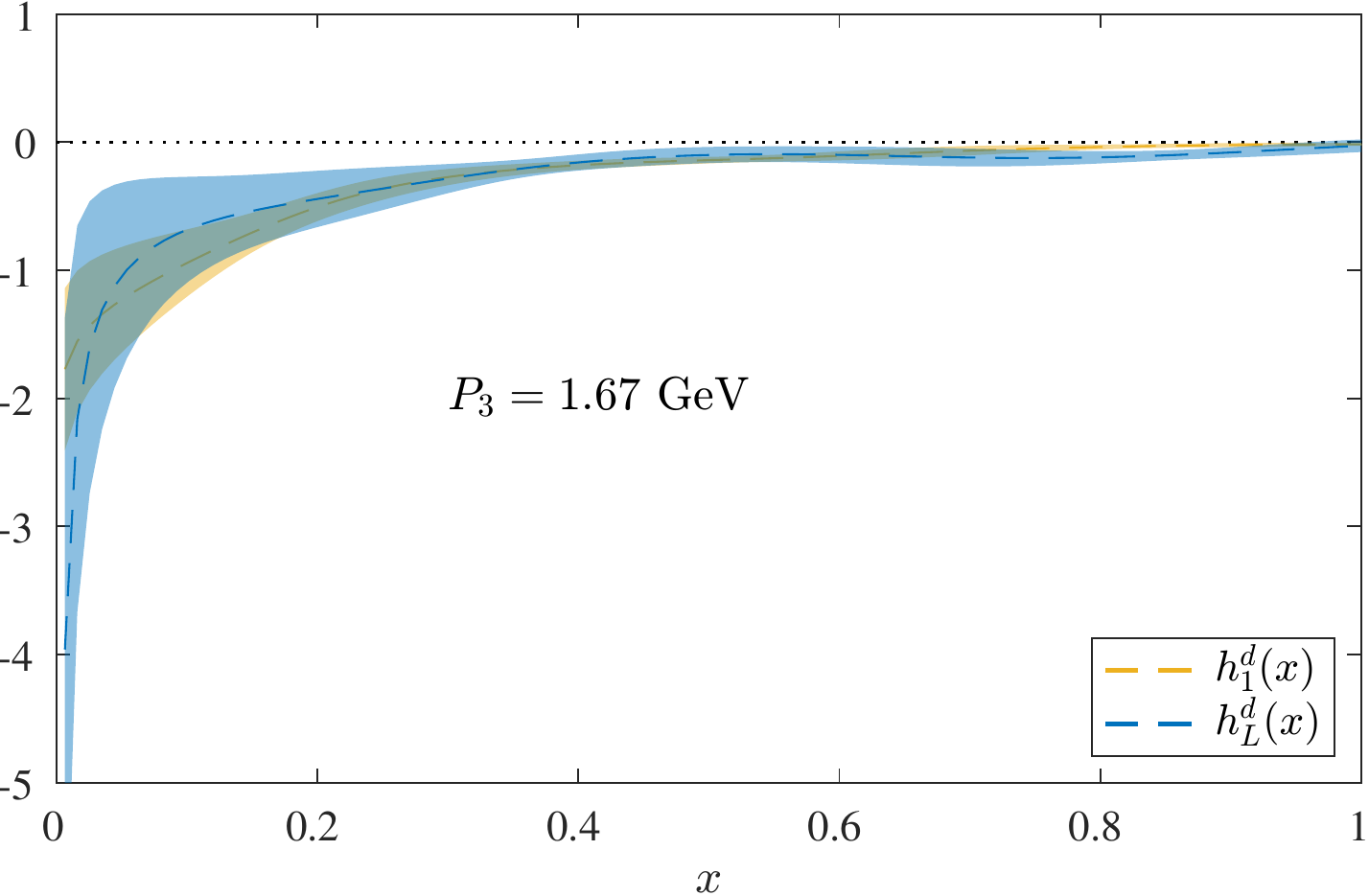} 
   \vspace*{-0.65cm}
\caption{$x$-dependence of $h_1$ (yellow) and $h_L$ (blue) for up-quarks and down-quarks in the left and right plot, respectively. Results are shown at the largest nucleon boost, $P_3=1.67$~GeV.} 
\label{h1_hL_u_d}
   \end{figure}

According to the finding of Ref.~\cite{Alexandrou:2021oih}, the disconnected contributions to $h_1^{u+d}(x)$ using the same ensemble as in this work are very small. We expect that the same applies for the operator entering $h_L(x)$. Therefore, we extract the up- and down-quark contributions using only the matrix elements extracted from the connected diagram. Another justification of the flavor decomposition is the fact that there is no gluon transversity, nor a twist-3 two-gluon matrix element for a longitudinally polarized target which could mix with $h_L(x)$~\cite{Mulders:2000sh}. In Fig.~\ref{h1_hL_u_d}, we show the individual quark contributions for $h_1(x)$ and $h_L(x)$ for $P_3{=}1.67$ GeV. We do not consider the antiquark region, which is very small and sensitive to systematic effects~\cite{Cichy:2018mum}. This comparison helps us to address qualitatively the role of the up and down quark in the proton, as well as their role in twist-2 and twist-3 PDFs. We find that the up quark is dominant in all regions of $x$, which is more pronounced for $x{<}0.5$. In the large-$x$ limit, $h_1^u(x)$ is about twice as large as $h_1^d(x)$. Similar conclusions are reached in the comparison between $h_L^u(x)$ and $h_L^d(x)$, with the former being dominant. Another observation is that the down quark has a similar contribution to $h_1(x)$ and $h_L(x)$ for all regions of $x$. In the case of the up quark, we find similar magnitude between $h_1^u(x)$ and $h_L^u(x)$ for $x{>}0.2$.

\begin{figure}[h!]
    \includegraphics[scale=0.52]{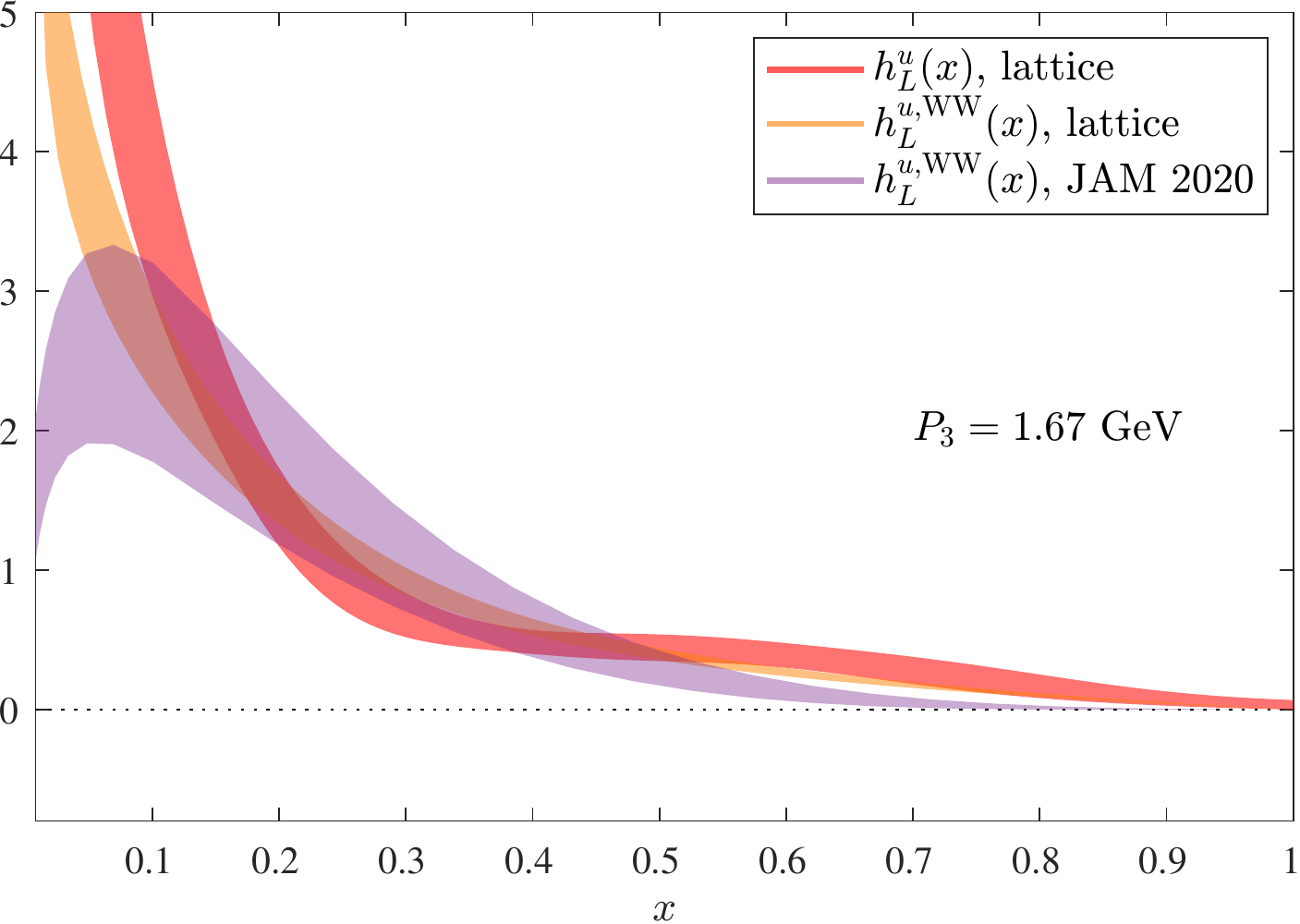}\hspace{0.1cm}
   \includegraphics[scale=0.52]{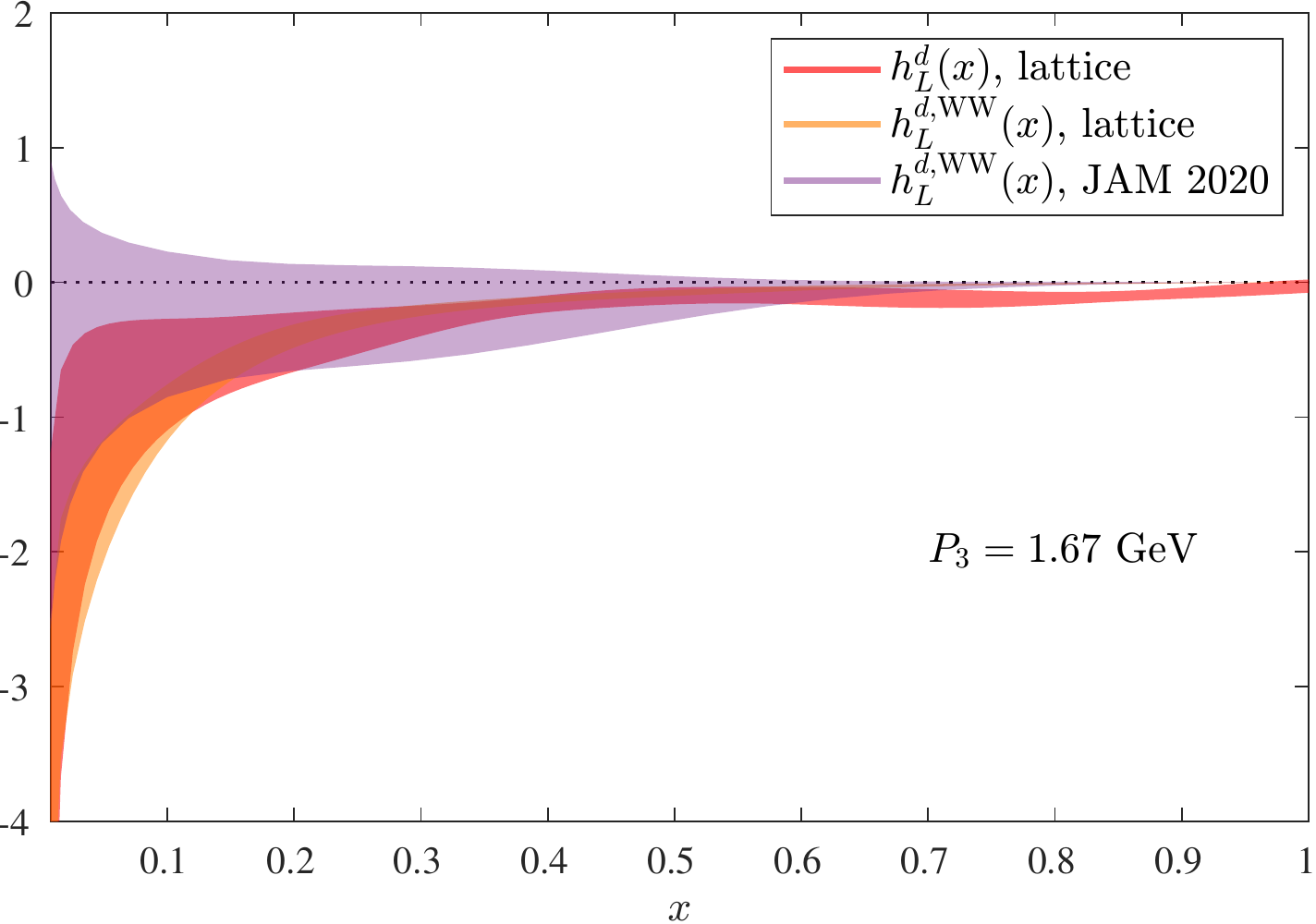} 
   \vspace*{-0.65cm}
\caption{Test of the WW approximation for up (left) and down (right) distributions for $P_3=1.67$~GeV. For the separate flavors, we show $h_L(x)$ (red) with $h_L^{\rm WW}(x)$ (orange)  extracted from lattice QCD within this work. Results for $h_L^{\rm WW}(x)$ from the JAM collaboration~\cite{Cammarota:2020qcw} (violet) are also included for comparison.}   
\label{WW_u_d}
   \end{figure}

It is interesting to test the WW approximation for each quark flavor, as shown in Fig.~\ref{WW_u_d}. While the isovector flavor combination for $h_L(x)$ and $h^{\rm WW}_L(x)$ at $P_3=1.67$ GeV show an agreement within uncertainties for $x<0.55$ (see, e.g., Fig.~\ref{fig:WW_approx}), here we find a discrepancy for this region for the up-quark. For the down-quark contribution, we find compatibility up to $x=0.7$. The JAM20 data~\cite{Cammarota:2020qcw} show agreement with the lattice data in the region between $x=0.15$ and $x=0.7$. We emphasize again that the data presented in this section neglect contributions from the disconnected diagram, which are expected to be within the reported uncertainties for $h_L(x)$~\cite{Alexandrou:2021oih}. 

\section{Summary}

In these proceedings, we summarize results on the first calculations of the twist-3 PDFs $g_T(x)$ and $h_L(x)$ for the proton. We employ the quasi-PDF approach, which connects lattice data to the light-cone PDFs using perturbative equations obtained in LaMET. The calculation is performed on an $N_f=2+1+1$ ensemble of twisted mass fermions with a clover term, corresponding to a pion mass of 260~MeV. The matrix elements are obtained for three values of the proton momentum, $P_3=0.83,\,1.25,\,1.67$ GeV, and the reconstruction of the quasi-PDFs uses the Backus-Gilbert method. The quasi-PDFs are matched to their light-cone counterparts using the matching equations obtained in one-loop perturbation theory~\cite{Bhattacharya:2020xlt,Bhattacharya:2020jfj}. These equations do not consider the mixing with quark-gluon correlations. Within the same setup, we also obtain the twist-2 $g_1(x)$ and $h_1(x)$, which are useful for qualitative comparison with their twist-3 counterparts. We find that the twist-3 PDFs are as large as the twist-2 PDFs.

The calculation of $g_1(x)$ allows us to explore the Burkhardt-Cottingham-type sum rule, which connects it to $g_T(x)$. We find that the sum rule is satisfied for the data of this work. For $h_L(x)$, we test the equivalent sum rule and find that it is both satisfied when compared to $h_1(x)$ and is, in addition, independent of the value of $P_3$. Another investigation is the exploration of the WW approximation, which also requires the twist-2 counterparts. 
For $h_L(x)$, we perform a flavor decomposition, as the disconnected contributions for $h_1(x)$ have been found to be small~\cite{Alexandrou:2021oih}. We find that $h_L^u(x)$ is positive and $h_L^d(x)$ negative, like is the case for $h_1^u(x)$ and $h_1^d(x)$.
For $x \in [0.1,0.5]$, there is little difference between $h_L(x)$ and $h_L^{\rm WW}(x)$ as obtained from our lattice data and the results from the JAM collaboration~\cite{Cammarota:2020qcw} (an exception is $h_L^u(x)$ in the region around $x \sim 0.3$). 

In the future, we plan to improve this calculation with an ensemble at the physical pion mass with larger volume and finer lattice spacing. We also plan to revisit the analysis for the matching along the lines of Refs.~\cite{Braun:2021gvv,Braun:2021aon}. Another improvement is the calculation of quark-gluon-quark correlations on the lattice, which requires a long-term dedicated program. We are currently extending this work to twist-3 GPDs, and preliminary results can be found in Ref.~\cite{Dodson_Lattice:2021}.

\vspace*{0.75cm}
\centerline{\textbf{Acknowledgements}}
\vspace*{0.2cm}

The work of S.B.~and A.M.~has been supported by the National Science Foundation under grant number PHY-2110472.  A.M.~has also been supported by the U.S. Department of Energy (DOE), Office of Nuclear Physics, within the framework of the TMD Topical Collaboration. K.C.\ is supported by the National Science Centre (Poland) grant SONATA BIS no.\ 2016/22/E/ST2/00013. M.C.~and A.S.~acknowledge financial support by the U.S. DOE, Office of Nuclear Physics, Early Career Award under Grant No.\ DE-SC0020405. F.S.\ was funded by by the NSFC and the Deutsche Forschungsgemeinschaft (DFG, German Research
Foundation) through the funds provided to the Sino-German Collaborative Research Center TRR110 (NSFC Grant No. 12070131001, DFG Project-ID 196253076 - TRR 110). Computations for this work were carried out in part on facilities of the USQCD Collaboration, which are funded by the Office of Science of the U.S. DOE. 
This research was supported in part by PLGrid Infrastructure (Prometheus supercomputer at AGH Cyfronet in Cracow).
Computations were also partially performed at the Poznan Supercomputing and Networking Center (Eagle supercomputer), the Interdisciplinary Centre for Mathematical and Computational Modelling of the Warsaw University (Okeanos supercomputer), and at the Academic Computer Centre in Gda\'nsk (Tryton supercomputer). The configurations have been generated on the KNL (A2) Partition of Marconi at CINECA, through the Prace project Pra13\_3304 ``SIMPHYS".
Inversions were performed using the DD-$\alpha$AMG solver with twisted mass support~\cite{Alexandrou:2016izb}.

%\bibliography{references}

\begin{thebibliography}{36}
\expandafter\ifx\csname natexlab\endcsname\relax\def\natexlab#1{#1}\fi
\expandafter\ifx\csname bibnamefont\endcsname\relax
  \def\bibnamefont#1{#1}\fi
\expandafter\ifx\csname bibfnamefont\endcsname\relax
  \def\bibfnamefont#1{#1}\fi
\expandafter\ifx\csname citenamefont\endcsname\relax
  \def\citenamefont#1{#1}\fi
\expandafter\ifx\csname url\endcsname\relax
  \def\url#1{\texttt{#1}}\fi
\expandafter\ifx\csname urlprefix\endcsname\relax\def\urlprefix{URL }\fi
\providecommand{\bibinfo}[2]{#2}
\providecommand{\eprint}[2][]{\url{#2}}

%\vspace*{-0.2cm}
%\bibitem[{\citenamefont{Collins and Soper}(1982)}]{Collins:1981uw}
%\bibinfo{author}{\bibfnamefont{J.~C.} \bibnamefont{Collins}} \bibnamefont{and}
%  \bibinfo{author}{\bibfnamefont{D.~E.} \bibnamefont{Soper}},
%  \bibinfo{journal}{Nucl.\ Phys.\ B} \textbf{\bibinfo{volume}{194}},
%  \bibinfo{pages}{445} (\bibinfo{year}{1982}).

\vspace*{-0.201cm}\bibitem[{\citenamefont{Jaffe}(1996)}]{Jaffe:1996zw}
\bibinfo{author}{\bibfnamefont{R.~L.} \bibnamefont{Jaffe}}, in
  \emph{\bibinfo{booktitle}{{The spin structure of the nucleon (1995)}}}
  (\bibinfo{year}{1996}), pp. \bibinfo{pages}{42--129},
  \eprint{hep-ph/9602236}.

\vspace*{-0.201cm}\bibitem[{\citenamefont{Balitsky and Braun}(1989)}]{Balitsky:1987bk}
\bibinfo{author}{\bibfnamefont{I.}~\bibnamefont{Balitsky}} \bibnamefont{and}
  \bibinfo{author}{\bibfnamefont{V.~M.} \bibnamefont{Braun}},
  \bibinfo{journal}{Nucl.\ Phys.\ B} \textbf{\bibinfo{volume}{311}},
  \bibinfo{pages}{541} (\bibinfo{year}{1989}).

%\vspace*{-0.201cm}\bibitem[{\citenamefont{Kanazawa et~al.}(2016)\citenamefont{Kanazawa, Koike,
%  Metz, Pitonyak, and Schlegel}}]{Kanazawa:2015ajw}
%\bibinfo{author}{\bibfnamefont{K.}~\bibnamefont{Kanazawa}},
%  \bibinfo{author}{\bibfnamefont{Y.}~\bibnamefont{Koike}},
%  \bibinfo{author}{\bibfnamefont{A.}~\bibnamefont{Metz}},
%  \bibinfo{author}{\bibfnamefont{D.}~\bibnamefont{Pitonyak}}, \bibnamefont{and}
%  \bibinfo{author}{\bibfnamefont{M.}~\bibnamefont{Schlegel}},
%  \bibinfo{journal}{Phys.\ Rev.\ D} \textbf{\bibinfo{volume}{93}},
%  \bibinfo{pages}{054024} (\bibinfo{year}{2016}), \eprint{1512.07233}.

%\vspace*{-0.201cm}\bibitem[{\citenamefont{Boer et~al.}(2003)\citenamefont{Boer, Mulders, and
%  Pijlman}}]{Boer:2003cm}
%\bibinfo{author}{\bibfnamefont{D.}~\bibnamefont{Boer}},
%  \bibinfo{author}{\bibfnamefont{P.~J.} \bibnamefont{Mulders}},
%  \bibnamefont{and} \bibinfo{author}{\bibfnamefont{F.}~\bibnamefont{Pijlman}},
%  \bibinfo{journal}{Nucl. Phys. B} \textbf{\bibinfo{volume}{667}},
%  \bibinfo{pages}{201} (\bibinfo{year}{2003}), \eprint{hep-ph/0303034}.

\vspace*{-0.201cm}\bibitem[{\citenamefont{Accardi et~al.}(2009)\citenamefont{Accardi, Bacchetta,
  Melnitchouk, and Schlegel}}]{Accardi:2009au}
\bibinfo{author}{\bibfnamefont{A.}~\bibnamefont{Accardi}},
  \bibinfo{author}{\bibfnamefont{A.}~\bibnamefont{Bacchetta}},
  \bibinfo{author}{\bibfnamefont{W.}~\bibnamefont{Melnitchouk}},
  \bibnamefont{and} \bibinfo{author}{\bibfnamefont{M.}~\bibnamefont{Schlegel}},
  \bibinfo{journal}{JHEP} \textbf{\bibinfo{volume}{11}}, \bibinfo{pages}{093}
  (\bibinfo{year}{2009}).%, \eprint{0907.2942}.

\vspace*{-0.201cm}\bibitem[{\citenamefont{Gamberg et~al.}(2018)\citenamefont{Gamberg, Metz,
  Pitonyak, and Prokudin}}]{Gamberg:2017jha}
\bibinfo{author}{\bibfnamefont{L.}~\bibnamefont{Gamberg}},
  \bibinfo{author}{\bibfnamefont{A.}~\bibnamefont{Metz}},
  \bibinfo{author}{\bibfnamefont{D.}~\bibnamefont{Pitonyak}}, \bibnamefont{and}
  \bibinfo{author}{\bibfnamefont{A.}~\bibnamefont{Prokudin}},
  \bibinfo{journal}{Phys. Lett. B} \textbf{\bibinfo{volume}{781}},
  \bibinfo{pages}{443} (\bibinfo{year}{2018}).%, \eprint{1712.08116}.

\vspace*{-0.201cm}\bibitem[{\citenamefont{Cammarota et~al.}(2020)\citenamefont{Cammarota,
  Gamberg, Kang, Miller, Pitonyak, Prokudin, Rogers, and
  Sato}}]{Cammarota:2020qcw}
\bibinfo{author}{\bibfnamefont{J.}~\bibnamefont{Cammarota}},
  \bibinfo{author}{\bibfnamefont{L.}~\bibnamefont{Gamberg}},
  \bibinfo{author}{\bibfnamefont{Z.-B.} \bibnamefont{Kang}},
  \bibinfo{author}{\bibfnamefont{J.~A.} \bibnamefont{Miller}},
  \bibinfo{author}{\bibfnamefont{D.}~\bibnamefont{Pitonyak}},
  \bibinfo{author}{\bibfnamefont{A.}~\bibnamefont{Prokudin}},
  \bibinfo{author}{\bibfnamefont{T.~C.} \bibnamefont{Rogers}},
  \bibnamefont{and} \bibinfo{author}{\bibfnamefont{N.}~\bibnamefont{Sato}}
  (\bibinfo{collaboration}{JAM}),
  \bibinfo{journal}{Phys. Rev. D} \textbf{\bibinfo{volume}{102}},
  \bibinfo{pages}{054002} (\bibinfo{year}{2020}), \eprint{2002.08384}.

\vspace*{-0.201cm}\bibitem[{\citenamefont{Burkardt}(2013)}]{Burkardt:2008ps}
\bibinfo{author}{\bibfnamefont{M.}~\bibnamefont{Burkardt}},
  \bibinfo{journal}{Phys. Rev.} \textbf{\bibinfo{volume}{D88}},
  \bibinfo{pages}{114502} (\bibinfo{year}{2013}), \eprint{0810.3589}.

%\cite{AbdulKhalek:2021gbh}
\vspace*{-0.201cm}
\bibitem{AbdulKhalek:2021gbh}
R.~Abdul Khalek,  \textit{et al.}
%``Science Requirements and Detector Concepts for the Electron-Ion Collider: EIC Yellow Report,''
\eprint{2103.05419}.
%144 citations counted in INSPIRE as of 24 Oct 2021

\vspace*{-0.201cm}\bibitem[{\citenamefont{Cichy and Constantinou}(2019)}]{Cichy:2018mum}
\bibinfo{author}{\bibfnamefont{K.}~\bibnamefont{Cichy}} \bibnamefont{and}
  \bibinfo{author}{\bibfnamefont{M.}~\bibnamefont{Constantinou}},
  \bibinfo{journal}{Adv. High Energy Phys.} \textbf{\bibinfo{volume}{2019}},
  \bibinfo{pages}{3036904} (\bibinfo{year}{2019}), \eprint{1811.07248}.

%\cite{Ji:2020ect}
\vspace*{-0.201cm} \bibitem{Ji:2020ect}
X.~Ji, Y.~S.~Liu, Y.~Liu, J.~H.~Zhang and Y.~Zhao,
%``Large-momentum effective theory,''
Rev. Mod. Phys. \textbf{93} (2021), 035005
\eprint{2004.03543}.
%75 citations counted in INSPIRE as of 01 Nov 2021

%\cite{Constantinou:2020pek}
\vspace*{-0.201cm}\bibitem{Constantinou:2020pek}
M.~Constantinou,
%``The x-dependence of hadronic parton distributions: A review on the progress of lattice QCD,''
Eur. Phys. J. A \textbf{57} (2021) no.2, 77
\eprint{2010.02445}.
%28 citations counted in INSPIRE as of 24 Oct 2021

%\cite{Cichy:2021lih}
\vspace*{-0.201cm}\bibitem{Cichy:2021lih}
K.~Cichy, PoS(LATTICE2021)017, 
%``Progress in $x$-dependent partonic distributions from lattice QCD,''
\eprint{2110.07440}.
%0 citations counted in INSPIRE as of 24 Oct 2021



%\vspace*{-0.201cm}\bibitem[{\citenamefont{Accardi et~al.}(2016)}]{Accardi:2012qut}
%\bibinfo{author}{\bibfnamefont{A.}~\bibnamefont{Accardi}} \bibnamefont{et~al.},
%  \bibinfo{journal}{Eur. Phys. J.} \textbf{\bibinfo{volume}{A52}},
%  \bibinfo{pages}{268} (\bibinfo{year}{2016}), \eprint{1212.1701}.

\vspace*{-0.201cm}\bibitem[{\citenamefont{Flay et~al.}(2016)}]{Flay:2016wie}
\bibinfo{author}{\bibfnamefont{D.}~\bibnamefont{Flay}} \bibnamefont{et~al.}
  (\bibinfo{collaboration}{Jefferson Lab Hall A}), \bibinfo{journal}{Phys.
  Rev.} \textbf{\bibinfo{volume}{D94}}, \bibinfo{pages}{052003}
  (\bibinfo{year}{2016}), \eprint{1603.03612}.

\vspace*{-0.201cm}\bibitem[{\citenamefont{Armstrong et~al.}(2019)}]{Armstrong:2018xgk}
\bibinfo{author}{\bibfnamefont{W.}~\bibnamefont{Armstrong}}
  \bibnamefont{et~al.} (\bibinfo{collaboration}{SANE}), \bibinfo{journal}{Phys.
  Rev. Lett.} \textbf{\bibinfo{volume}{122}}, \bibinfo{pages}{022002}
  (\bibinfo{year}{2019}), \eprint{1805.08835}.

\vspace*{-0.201cm}\bibitem[{\citenamefont{Jaffe and Ji}(1991)}]{Jaffe:1991kp}
\bibinfo{author}{\bibfnamefont{R.~L.} \bibnamefont{Jaffe}} \bibnamefont{and}
  \bibinfo{author}{\bibfnamefont{X.-D.} \bibnamefont{Ji}},
  \bibinfo{journal}{Phys. Rev. Lett.} \textbf{\bibinfo{volume}{67}},
  \bibinfo{pages}{552} (\bibinfo{year}{1991}).

\vspace*{-0.201cm}\bibitem[{\citenamefont{Jaffe and Ji}(1992)}]{Jaffe:1991ra}
\bibinfo{author}{\bibfnamefont{R.~L.} \bibnamefont{Jaffe}} \bibnamefont{and}
  \bibinfo{author}{\bibfnamefont{X.-D.} \bibnamefont{Ji}},
  \bibinfo{journal}{Nucl. Phys. B} \textbf{\bibinfo{volume}{375}},
  \bibinfo{pages}{527} (\bibinfo{year}{1992}).

%\vspace*{-0.201cm}\bibitem[{\citenamefont{Koike et~al.}(2008)\citenamefont{Koike, Tanaka, and
%  Yoshida}}]{Koike:2008du}
%\bibinfo{author}{\bibfnamefont{Y.}~\bibnamefont{Koike}},
%  \bibinfo{author}{\bibfnamefont{K.}~\bibnamefont{Tanaka}}, \bibnamefont{and}
%  \bibinfo{author}{\bibfnamefont{S.}~\bibnamefont{Yoshida}},
%  \bibinfo{journal}{Phys. Lett. B} \textbf{\bibinfo{volume}{668}},
%  \bibinfo{pages}{286} (\bibinfo{year}{2008}), \eprint{0805.2289}.

\vspace*{-0.201cm}\bibitem[{\citenamefont{Koike et~al.}(2016)\citenamefont{Koike, Pitonyak, and
  Yoshida}}]{Koike:2016ura}
\bibinfo{author}{\bibfnamefont{Y.}~\bibnamefont{Koike}},
  \bibinfo{author}{\bibfnamefont{D.}~\bibnamefont{Pitonyak}}, \bibnamefont{and}
  \bibinfo{author}{\bibfnamefont{S.}~\bibnamefont{Yoshida}},
  \bibinfo{journal}{Phys. Lett. B} \textbf{\bibinfo{volume}{759}},
  \bibinfo{pages}{75} (\bibinfo{year}{2016}), \eprint{1603.07908}.

\vspace*{-0.201cm}\bibitem[{\citenamefont{Bhattacharya
  et~al.}(2020{\natexlab{a}})\citenamefont{Bhattacharya, Cichy, Constantinou,
  Metz, Scapellato, and Steffens}}]{Bhattacharya:2020cen}
\bibinfo{author}{\bibfnamefont{S.}~\bibnamefont{Bhattacharya}},
  \bibinfo{author}{\bibfnamefont{K.}~\bibnamefont{Cichy}},
  \bibinfo{author}{\bibfnamefont{M.}~\bibnamefont{Constantinou}},
  \bibinfo{author}{\bibfnamefont{A.}~\bibnamefont{Metz}},
  \bibinfo{author}{\bibfnamefont{A.}~\bibnamefont{Scapellato}},
  \bibnamefont{and} \bibinfo{author}{\bibfnamefont{F.}~\bibnamefont{Steffens}},
  \bibinfo{journal}{Phys. Rev. D} \textbf{\bibinfo{volume}{102}},
  \bibinfo{pages}{111501} (\bibinfo{year}{2020}{\natexlab{a}}),
  \eprint{2004.04130}.

\vspace*{-0.201cm}\bibitem[{\citenamefont{Bhattacharya et~al.}(2021)\citenamefont{Bhattacharya,
  Cichy, Constantinou, Metz, Scapellato, and Steffens}}]{Bhattacharya:2021moj}
\bibinfo{author}{\bibfnamefont{S.}~\bibnamefont{Bhattacharya}},
  \bibinfo{author}{\bibfnamefont{K.}~\bibnamefont{Cichy}},
  \bibinfo{author}{\bibfnamefont{M.}~\bibnamefont{Constantinou}},
  \bibinfo{author}{\bibfnamefont{A.}~\bibnamefont{Metz}},
  \bibinfo{author}{\bibfnamefont{A.}~\bibnamefont{Scapellato}},
  \bibnamefont{and} \bibinfo{author}{\bibfnamefont{F.}~\bibnamefont{Steffens}}
  (\bibinfo{year}{2021}), \eprint{2107.02574}.

\vspace*{-0.201cm}\bibitem[{\citenamefont{Ji}(2013)}]{Ji:2013dva}
\bibinfo{author}{\bibfnamefont{X.}~\bibnamefont{Ji}}, \bibinfo{journal}{Phys.
  Rev. Lett.} \textbf{\bibinfo{volume}{110}}, \bibinfo{pages}{262002}
  (\bibinfo{year}{2013}), \eprint{1305.1539}.

\vspace*{-0.201cm}\bibitem[{\citenamefont{Ji}(2014)}]{Ji:2014gla}
\bibinfo{author}{\bibfnamefont{X.}~\bibnamefont{Ji}}, \bibinfo{journal}{Sci.
  China Phys. Mech. Astron.} \textbf{\bibinfo{volume}{57}},
  \bibinfo{pages}{1407} (\bibinfo{year}{2014}), \eprint{1404.6680}.

\vspace*{-0.201cm}\bibitem[{\citenamefont{Bhattacharya
  et~al.}(2020{\natexlab{b}})\citenamefont{Bhattacharya, Cichy, Constantinou,
  Metz, Scapellato, and Steffens}}]{Bhattacharya:2020xlt}
\bibinfo{author}{\bibfnamefont{S.}~\bibnamefont{Bhattacharya}},
  \bibinfo{author}{\bibfnamefont{K.}~\bibnamefont{Cichy}},
  \bibinfo{author}{\bibfnamefont{M.}~\bibnamefont{Constantinou}},
  \bibinfo{author}{\bibfnamefont{A.}~\bibnamefont{Metz}},
  \bibinfo{author}{\bibfnamefont{A.}~\bibnamefont{Scapellato}},
  \bibnamefont{and} \bibinfo{author}{\bibfnamefont{F.}~\bibnamefont{Steffens}},
  \bibinfo{journal}{Phys. Rev. D} \textbf{\bibinfo{volume}{102}},
  \bibinfo{pages}{034005} (\bibinfo{year}{2020}{\natexlab{b}}),
  \eprint{2005.10939}.

\vspace*{-0.201cm}\bibitem[{\citenamefont{Bhattacharya
  et~al.}(2020{\natexlab{c}})\citenamefont{Bhattacharya, Cichy, Constantinou,
  Metz, Scapellato, and Steffens}}]{Bhattacharya:2020jfj}
\bibinfo{author}{\bibfnamefont{S.}~\bibnamefont{Bhattacharya}},
  \bibinfo{author}{\bibfnamefont{K.}~\bibnamefont{Cichy}},
  \bibinfo{author}{\bibfnamefont{M.}~\bibnamefont{Constantinou}},
  \bibinfo{author}{\bibfnamefont{A.}~\bibnamefont{Metz}},
  \bibinfo{author}{\bibfnamefont{A.}~\bibnamefont{Scapellato}},
  \bibnamefont{and} \bibinfo{author}{\bibfnamefont{F.}~\bibnamefont{Steffens}},
  \bibinfo{journal}{Phys. Rev. D} \textbf{\bibinfo{volume}{102}},
  \bibinfo{pages}{114025} (\bibinfo{year}{2020}{\natexlab{c}}),
  \eprint{2006.12347}.


%\cite{Braun:2021gvv}
\vspace*{-0.201cm}\bibitem{Braun:2021gvv}
V.~M.~Braun, Y.~Ji and A.~Vladimirov,
%``QCD factorization for chiral-odd parton quasi- and pseudo-distributions,''
JHEP \textbf{10} (2021), 087,
\eprint{2108.03065}.
%1 citations counted in INSPIRE as of 24 Oct 2021


\vspace*{-0.201cm}\bibitem[{\citenamefont{Braun et~al.}(2021)\citenamefont{Braun, Ji, and
  Vladimirov}}]{Braun:2021aon}
\bibinfo{author}{\bibfnamefont{V.~M.} \bibnamefont{Braun}},
  \bibinfo{author}{\bibfnamefont{Y.}~\bibnamefont{Ji}}, \bibnamefont{and}
  \bibinfo{author}{\bibfnamefont{A.}~\bibnamefont{Vladimirov}},
  \bibinfo{journal}{JHEP} \textbf{\bibinfo{volume}{05}}, \bibinfo{pages}{086}
  (\bibinfo{year}{2021}), \eprint{2103.12105}.

\vspace*{-0.201cm}\bibitem[{\citenamefont{Alexandrou
  et~al.}(2021{\natexlab{a}})}]{Alexandrou:2021gqw}
\bibinfo{author}{\bibfnamefont{C.}~\bibnamefont{Alexandrou}}
  \bibnamefont{et~al.} (\bibinfo{year}{2021}{\natexlab{a}}),
  \eprint{2104.13408}.

\vspace*{-0.201cm}\bibitem[{\citenamefont{Burkhardt and Cottingham}(1970)}]{Burkhardt:1970ti}
\bibinfo{author}{\bibfnamefont{H.}~\bibnamefont{Burkhardt}} \bibnamefont{and}
  \bibinfo{author}{\bibfnamefont{W.~N.} \bibnamefont{Cottingham}},
  \bibinfo{journal}{Annals Phys.} \textbf{\bibinfo{volume}{56}},
  \bibinfo{pages}{453} (\bibinfo{year}{1970}).

\vspace*{-0.201cm}\bibitem[{\citenamefont{Burkardt}(1995)}]{Burkardt:1995ts}
\bibinfo{author}{\bibfnamefont{M.}~\bibnamefont{Burkardt}},
  \bibinfo{journal}{Phys. Rev. D} \textbf{\bibinfo{volume}{52}},
  \bibinfo{pages}{3841} (\bibinfo{year}{1995}), \eprint{hep-ph/9505226}.

\vspace*{-0.201cm}\bibitem[{\citenamefont{Bhattacharya and Metz}(2021)}]{Bhattacharya:2021boh}
\bibinfo{author}{\bibfnamefont{S.}~\bibnamefont{Bhattacharya}}
  \bibnamefont{and} \bibinfo{author}{\bibfnamefont{A.}~\bibnamefont{Metz}}
  (\bibinfo{year}{2021}), \eprint{2105.07282}.

\vspace*{-0.201cm}\bibitem[{\citenamefont{Wandzura and Wilczek}(1977)}]{Wandzura:1977qf}
\bibinfo{author}{\bibfnamefont{S.}~\bibnamefont{Wandzura}} \bibnamefont{and}
  \bibinfo{author}{\bibfnamefont{F.}~\bibnamefont{Wilczek}},
  \bibinfo{journal}{Phys. Lett.} \textbf{\bibinfo{volume}{72B}},
  \bibinfo{pages}{195} (\bibinfo{year}{1977}).

\vspace*{-0.201cm}\bibitem[{\citenamefont{Dressler and Polyakov}(2000)}]{Dressler:1999hc}
\bibinfo{author}{\bibfnamefont{B.}~\bibnamefont{Dressler}} \bibnamefont{and}
  \bibinfo{author}{\bibfnamefont{M.~V.} \bibnamefont{Polyakov}},
  \bibinfo{journal}{Phys. Rev. D} \textbf{\bibinfo{volume}{61}},
  \bibinfo{pages}{097501} (\bibinfo{year}{2000}), \eprint{hep-ph/9912376}.

\vspace*{-0.201cm}\bibitem[{\citenamefont{Nocera et~al.}(2014)\citenamefont{Nocera, Ball, Forte,
  Ridolfi, and Rojo}}]{Nocera:2014gqa}
\bibinfo{author}{\bibfnamefont{E.~R.} \bibnamefont{Nocera}},
  \bibinfo{author}{\bibfnamefont{R.~D.} \bibnamefont{Ball}},
  \bibinfo{author}{\bibfnamefont{S.}~\bibnamefont{Forte}},
  \bibinfo{author}{\bibfnamefont{G.}~\bibnamefont{Ridolfi}}, \bibnamefont{and}
  \bibinfo{author}{\bibfnamefont{J.}~\bibnamefont{Rojo}}
  , \bibinfo{journal}{Nucl. Phys.}
  \textbf{\bibinfo{volume}{B887}}, \bibinfo{pages}{276} (\bibinfo{year}{2014}).

\vspace*{-0.201cm}\bibitem[{\citenamefont{Ethier et~al.}(2017)\citenamefont{Ethier, Sato, and
  Melnitchouk}}]{Ethier:2017zbq}
\bibinfo{author}{\bibfnamefont{J.~J.} \bibnamefont{Ethier}},
  \bibinfo{author}{\bibfnamefont{N.}~\bibnamefont{Sato}}, \bibnamefont{and}
  \bibinfo{author}{\bibfnamefont{W.}~\bibnamefont{Melnitchouk}},
  \bibinfo{journal}{Phys. Rev. Lett.} \textbf{\bibinfo{volume}{119}},
  \bibinfo{pages}{132001} (\bibinfo{year}{2017}), \eprint{1705.05889}.

\vspace*{-0.201cm}\bibitem[{\citenamefont{Alexandrou
  et~al.}(2021{\natexlab{b}})\citenamefont{Alexandrou, Constantinou,
  Hadjiyiannakou, Jansen, and Manigrasso}}]{Alexandrou:2021oih}
\bibinfo{author}{\bibfnamefont{C.}~\bibnamefont{Alexandrou}},
  \bibinfo{author}{\bibfnamefont{M.}~\bibnamefont{Constantinou}},
  \bibinfo{author}{\bibfnamefont{K.}~\bibnamefont{Hadjiyiannakou}},
  \bibinfo{author}{\bibfnamefont{K.}~\bibnamefont{Jansen}}, \bibnamefont{and}
  \bibinfo{author}{\bibfnamefont{F.}~\bibnamefont{Manigrasso}},
  Phys. Rev. D \textbf{104} (2021) no.5, 054503,
  (\bibinfo{year}{2021}{\natexlab{b}}), \eprint{2106.16065}.

\vspace*{-0.201cm}\bibitem[{\citenamefont{Mulders and Rodrigues}(2001)}]{Mulders:2000sh}
\bibinfo{author}{\bibfnamefont{P.~J.} \bibnamefont{Mulders}} \bibnamefont{and}
  \bibinfo{author}{\bibfnamefont{J.}~\bibnamefont{Rodrigues}},
  \bibinfo{journal}{Phys. Rev. D} \textbf{\bibinfo{volume}{63}},
  \bibinfo{pages}{094021} (\bibinfo{year}{2001}), \eprint{hep-ph/0009343}.


%\vspace*{-0.201cm}\bibitem[{\citenamefont{Frommer et~al.}(2014)\citenamefont{Frommer, Kahl,
%  Krieg, Leder, and Rottmann}}]{Frommer:2013fsa}
%\bibinfo{author}{\bibfnamefont{A.}~\bibnamefont{Frommer}},
%%  \bibinfo{author}{\bibfnamefont{K.}~\bibnamefont{Kahl}},
%  \bibinfo{author}{\bibfnamefont{S.}~\bibnamefont{Krieg}},
%  \bibinfo{author}{\bibfnamefont{B.}~\bibnamefont{Leder}}, \bibnamefont{and}
%  \bibinfo{author}{\bibfnamefont{M.}~\bibnamefont{Rottmann}},
%  \bibinfo{journal}{SIAM J. Sci. Comput.} \textbf{\bibinfo{volume}{36}},
%  \bibinfo{pages}{A1581} (\bibinfo{year}{2014}), \eprint{1303.1377}.

\vspace*{-0.201cm}\bibitem[{\citenamefont{Alexandrou et~al.}(2016)\citenamefont{Alexandrou,
  Bacchio, Finkenrath, Frommer, Kahl, and Rottmann}}]{Alexandrou:2016izb}
\bibinfo{author}{\bibfnamefont{C.}~\bibnamefont{Alexandrou}},
  \bibinfo{author}{\bibfnamefont{S.}~\bibnamefont{Bacchio}},
  \bibinfo{author}{\bibfnamefont{J.}~\bibnamefont{Finkenrath}},
  \bibinfo{author}{\bibfnamefont{A.}~\bibnamefont{Frommer}},
  \bibinfo{author}{\bibfnamefont{K.}~\bibnamefont{Kahl}}, \bibnamefont{and}
  \bibinfo{author}{\bibfnamefont{M.}~\bibnamefont{Rottmann}},
  \bibinfo{journal}{Phys. Rev. D} \textbf{\bibinfo{volume}{94}},
  \bibinfo{pages}{114509} (\bibinfo{year}{2016}), \eprint{1610.02370}.
  
  
\vspace*{-0.201cm}\bibitem[{\citenamefont{Bhattacharya
  et~al.}(2020{\natexlab{a}})\citenamefont{Bhattacharya, Cichy, Constantinou, Dodson,
  Metz, Scapellato, and Steffens}}]{Dodson_Lattice:2021}
\bibinfo{author}{\bibfnamefont{S.}~\bibnamefont{Bhattacharya}},
  \bibinfo{author}{\bibfnamefont{K.}~\bibnamefont{Cichy}},
  \bibinfo{author}{\bibfnamefont{M.}~\bibnamefont{Constantinou}},
  \bibinfo{author}{\bibfnamefont{J.}~\bibnamefont{Dodson}},
  \bibinfo{author}{\bibfnamefont{A.}~\bibnamefont{Metz}},
  \bibinfo{author}{\bibfnamefont{A.}~\bibnamefont{Scapellato}},
  \bibnamefont{and} \bibinfo{author}{\bibfnamefont{F.}~\bibnamefont{Steffens}},
  \bibinfo{journal}{PoS(LATTICE2021)054}.
  

\end{thebibliography}

\end{document}